\documentclass[sigconf,9pt]{acmart}
\AtBeginDocument{%
  }

\copyrightyear{2026}
\acmYear{2026}
\setcopyright{cc}
\setcctype{by}
\acmConference[DAC '26]{63rd ACM/IEEE Design Automation Conference}{July 26--29, 2026}{Long Beach, CA, USA}
\acmBooktitle{63rd ACM/IEEE Design Automation Conference (DAC '26), July 26--29, 2026, Long Beach, CA, USA}
\acmDOI{10.1145/3770743.3804087}
\acmISBN{979-8-4007-2254-7/2026/07}

\usepackage{graphicx} 
\usepackage{braket}
\usepackage{subfigure}

\usepackage{amsmath,amsthm,amsfonts}

\usepackage{amssymb}

\usepackage{xcolor}

\usepackage{tikz}
\definecolor{custompurple}{rgb}{0.5, 0.1, 0.5}
\usepackage{circuitikz}
\usetikzlibrary{quantikz}
\usepackage{setspace} 
\usepackage{circledsteps}

\pgfkeys{/csteps/inner ysep=10pt}
\usepackage{yquant}

\usepackage{ifthen}
\useyquantlanguage{groups}
\usepackage{pgfplots}
\usetikzlibrary{quotes}
\usepackage{pifont}
\pgfplotsset{compat=1.18}

\usepackage[normalem]{ulem}

\usepackage[htt]{hyphenat}

\usepackage{comment}
\newtheorem{thm}{Theorem}

\usepackage{algorithm}
\usepackage{algorithmicx}
\usepackage{algpseudocode}

\makeatletter
\def\@fnsymbol#1{%
  \ifcase#1\or
    \mbox{}
  \else
    \@ctrerr
  \fi
}
\makeatother

\begin{document}

\title{Fast Template Matching for Quantum Circuits Using Hypergraph}


\author{Xin Hong$^1$, Jintao Yu$^2$, Runhong He$^1$, Shenggang Ying$^{1*}$, Sanjiang Li$^3$, Mingsheng Ying$^3$}
\authornote{*Corresponding author: yingsg@ios.ac.cn}
\affiliation{%
  \institution{$^1$Key Laboratory of System Software (Chinese Academy of Sciences), Institute of Software, Chinese Academy of Sciences}
    \state{Beijing}
  \country{China}
  }
\affiliation{%
  \institution{$^2$Arclight Quantum Computing Inc., Beijing, China}
  \country{}
  }
\affiliation{%
  \institution{$^3$University of Technology Sydney, Sydney, Australia}
  \country{}
  }

\renewcommand{\shortauthors}{Hong et al.}

\begin{abstract}



Quantum circuit optimization is a key step toward the efficient execution of quantum algorithms. Template matching has emerged as one of the dominant approaches for simplifying quantum circuits, yet it confronts the intrinsic challenge of accommodating gate commutativity. The state-of-the-art template-matching algorithm relies on a directed acyclic graph (DAG) representation. While this DAG-based technique handles gate commutativity satisfactorily, it fails to preserve the local connectivity inherent to quantum circuits, thereby leading to relatively high matching complexity.

In this paper, we introduce a \emph{hypergraph representation} (HG), a commutativity\hyp{}aware representation that collapses commuting gates into a single \emph{super-node} while \emph{retaining all local connectivity}.
This enables matches to be extended \emph{locally} without revisiting the rest of the circuit, with incremental updates limited to the immediate neighborhood.
Experimental results demonstrate that our HG matcher achieves \textbf{11--606$\times$ speed-up} over the DAG-based implementation in Qiskit (Iten et al., 2022) on both random and arithmetic benchmarks, while maintaining the same optimization quality.
The acceleration increases with circuit size, confirming that preserving locality and connectivity is the key to scalable quantum circuit optimisation.

\end{abstract}

\ccsdesc[500]{Hardware~Quantum technologies}

 \keywords{quantum circuits, template match, hypergraph}

\maketitle

\section{Introduction}

Quantum algorithms promise exponential speed-ups over their classical counterparts, but only if the underlying quantum circuits can be compiled into hardware-feasible, low-depth implementations. Near-term Noisy Intermediate-Scale Quantum (NISQ) devices impose stringent constraints on qubit count, connectivity, and coherence time~\cite{preskill2018quantum,bharti2022noisy,arute2019quantum}, 
making aggressive circuit optimization an important enabler for the effective implementation of advanced quantum algorithms
~\cite{cerezo2021variational,bravyi2020quantum}.

Among existing optimization strategies, \emph{template-based simplification} has emerged as one of the most powerful~\cite{nam2018automated,amy2014polynomial,fischbach2025review,gao2023quantum,amy2019t}.
A template is a small, functionally equivalent circuit fragment that realizes the identity; whenever a structural match is detected, the expensive sub-circuit is replaced by the cheaper \emph{inverse remainder}, yielding a net reduction in gate count~\cite{iten_exact_2022}.
The central challenge, however, is \emph{not} merely finding isomorphic sub-circuits; it is doing so while respecting the fact that many quantum gates commute~\cite{maslov2008quantum,cole2021commuting,xu2025optimizing}.
Ignoring commutativity misses valuable simplification opportunities, whereas naively enumerating all permutations quickly becomes intractable.

The current leading solution represents the circuit as a directed acyclic graph (DAG), in which vertices are gates and edges indicate gate dependencies and non-commutativity \cite{iten_exact_2022}.
By topologically comparing and matching DAGs, the algorithm identifies candidate matches in quasi-quadratic time.
This approach elegantly sidesteps the commutativity explosion, but it comes at a steep price: \emph{edges between commuting gates are discarded}, reducing the circuit to a set of mutually independent subgraphs.
Consequently,
\vspace{-3pt}
\begin{enumerate}
  \item \textbf{Connectivity is lost}---two gates that share a qubit but commute are treated as if they were on different qubits of the circuit; and
 \item \textbf{Locality is lost}---sometimes the matcher needs to search globally to find the next matchable gates, even if some of the gates have already been matched.
\end{enumerate}
\vspace{-3pt}

In this work, we ask whether it is possible to retain the DAG’s commutativity-aware efficiency without sacrificing circuit structure detail. We answer the question affirmatively with a new hypergraph representation (HG). HG is a commutativity-aware representation that keeps the \emph{efficiency} of the DAG approach while \emph{preserving} the local structure of the circuit.
Instead of deleting edges between commuting gates, we \emph{collapse them into a single super-node} that stores the entire algebraic equivalence class.
Edges encoding true gate dependencies are retained, so the HG remains a \emph{connected} object.
As a result, after a partial match is found, the algorithm only needs to explore the immediate neighbourhood of the matched super-node; there is no need to restart a global search.

Experimental results on random circuits, fixed arithmetic benchmarks, and scalability tests show that HG achieves \textbf{11--606$\times$ speed-up} over the state-of-the-art DAG-based matcher in Qiskit \cite{iten_exact_2022} while maintaining the same optimization quality.
The acceleration grows with circuit size, confirming that preserving locality and connectivity is not merely a theoretical convenience---it is the key to scalable quantum circuit optimisation.

The remainder of the paper is organized as follows. Section \ref{sec:background} introduces basic concepts of quantum circuits. Section \ref{sec:SG} introduces the hypergraph representation and proves that it preserves both commutativity and canonicity. Section \ref{sec:match} describes the subgraph-matching algorithm. An example is given in section \ref{sec:example}. Section \ref{sec:experiments} presents experimental results, and section \ref{sec:conclusion} concludes the paper.

\section{Background}\label{sec:background}

\subsection{Qubits and Quantum gates}
A qubit is a two-level quantum system whose state is written as
\[
|\psi\rangle=\alpha|0\rangle+\beta|1\rangle,\qquad \alpha,\beta\in\mathbb{C},\quad |\alpha|^{2}+|\beta|^{2}=1.
\]
An $n$-qubit register lives in $(\mathbb{C}^{2})^{\otimes n}$ and its state is described by $2^{n}$ complex amplitudes.

Computations are realized by unitary operators called quantum gates. Single-qubit examples are the Pauli-$X$ gate and the Pauli-$Z$ gate: 
\[
X=\begin{pmatrix}0 & 1\\ 1 & 0\end{pmatrix},
\quad
Z=\begin{pmatrix}1 & 0\\ 0 & -1\end{pmatrix}.
\]
Two-qubit gates such as $CX$
\[
CX=|0\rangle\langle 0|\otimes I+|1\rangle\langle 1|\otimes X
\]
entangle qubits. A gate set is \emph{universal} if any unitary can be approximated to arbitrary precision by products of gates from the gate set. 
In this paper, we focus on quantum gates that have only one target qubit but may involve zero, one, or multiple control qubits. We refer to these as multi-controlled single-target gates.

Eigen-directions (generalized eigenbases) are used to decide commutativity. Let $U$ be a quantum gate, we say $\ket{\psi}$ is an \textbf{eigen-direction} of $U$ if $U\ket{\psi}=\lambda\ket{\psi}$ for some $\lambda \in \mathbb{C}$ and $\lambda \neq 0$. Then, the eigen-directions of $X$ are $\{\ket{+}, \ket{-}\}$, and the eigen-directions of $Z$ are $\{\ket{0}, \ket{1}\}$. For controlled quantum gates, we consider the eigen-direction per qubit. Specifically, the eigen-directions of any control qubits are $\{\ket{0}, \ket{1}\}$ and the eigen-directions on the target qubit depend on the quantum gate applied to the target qubit. For example, for $CX$, the eigen-directions on target qubit are $\{\ket{+}, \ket{-}\}$.

\subsection{Quantum Circuits}
A quantum circuit is an ordered sequence $\mathcal{C}=(g_{1},\dots,g_{m})$ where each $g_{i}$ is a gate acting on specified qubit wires. The circuit defines a unitary
\[
U_{\mathcal{C}}=g_{m}\dots g_{2}g_{1}.
\]
The \emph{width} of $C$ is the number of qubits $n$, sometimes written $n_C$ for clarification; the \emph{depth} of $C$ is the longest path from input to output; the \emph{size} of $C$, denoted as $|C|$, is the total number of gates $m$. 

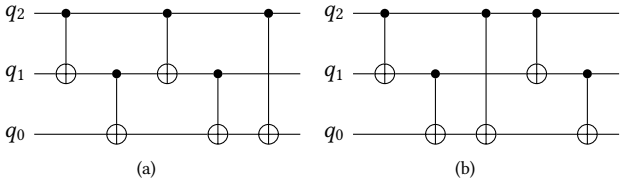
\begin{figure}[htbp]
\centering
\vspace{-10pt}
\subfigure[]{
\begin{tikzpicture}
  \begin{yquant}[register/minimum height=7mm, operator/separation=1.7mm, control style={radius=1.7pt}, subcircuit box style={dashed}]
    qubit {$q_2$} q2;
    qubit {$q_1$} q1;
    qubit {$q_0$} q0;

    cnot q1 | q2;
    cnot q0 | q1;
    cnot q1 | q2;
    cnot q0 | q1;
    cnot q0 | q2;
    
  \end{yquant}
  
\end{tikzpicture}
}
\subfigure[]{
\begin{tikzpicture}
  \begin{yquant}[register/minimum height=7mm, operator/separation=1.7mm, control style={radius=1.7pt}, subcircuit box style={dashed}]
    qubit {$q_2$} q2;
    qubit {$q_1$} q1;
    qubit {$q_0$} q0;

    cnot q1 | q2;
    cnot q0 | q1;
    cnot q0 | q2;
    cnot q1 | q2;
    cnot q0 | q1;
    
  \end{yquant}
\end{tikzpicture}

}
\vspace{-10pt}
\caption{Two equivalent quantum circuits consisting of 3 qubits and 5 gates. Both circuits implement the identity matrix. 
}
\vspace{-15pt}
\label{fig:qc1}
\end{figure}

\subsection{Gate commutativity and Circuit equivalence}
Two gates $g_{i}$ and $g_{j}$ \emph{commute} if $g_{i}g_{j}=g_{j}g_{i}$.
Commutativity depends on qubit overlap: gates on disjoint wires always commute, whereas gates sharing at least one wire may not. Specifically, we say that two gates are \textbf{commuting on qubit $q$} if they have the same eigen-directions on the qubit. 
Commutativity can lead to the emergence of many equivalent but distinct circuits. 

Circuits $\mathcal{C}_{1}$ and $\mathcal{C}_{2}$ are \emph{equivalent} if they realize the same unitary:
\[
\mathcal{C}_{1}\equiv\mathcal{C}_{2}\quad\Longleftrightarrow\quad U_{\mathcal{C}_{1}}=U_{\mathcal{C}_{2}}.
\]
Fig.~\ref{fig:qc1} shows two equivalent circuits that implement the identity matrix. 
In particular, we define two circuits to be \emph{equivalent under gate commutation}, or \emph{commutationally equivalent}, denoted as
\begin{equation}
C_1\equiv_{\text{comm}}C_2,
\end{equation} 
if one can be obtained from the other by sequential swapping of commutable gates.


    

  

\subsection{Template Matching}

A \emph{template }is an identity circuit
\[
\mathcal{T}=(g_{1},\cdots, g_{k}),\qquad U_{\mathcal{T}}=I.
\]
Given a target circuit $\mathcal{C}$, a compiler searches for a sub-sequence
\[
\mathcal{M}=(h_{p_{1}},\cdots, h_{p_{m}})\subseteq\mathcal{C}
\]
that is structurally equivalent to $m>k/2$ consecutive gates of $\mathcal{T}$ (modulo commutation), for example $(g_{p_{1}},\cdots, g_{p_{m}})$.  
The matched segment $\mathcal{M}$ is replaced by the \emph{inverse remainder}
\[
\mathcal{T}_{\text{rem}}^{-1}=(g_{p_{1}-1}^{-1},\cdots ,g_{1}^{-1},g_{k}^{-1},\cdots, g_{p_{m}+1}^{-1}).
\]
which contains fewer than $m$ gates, yielding a net reduction while preserving the overall unitary because $U_{\mathcal{M}}=U_{\mathcal{T}_{\text{rem}}^{-1}}$.
The algorithmic challenge is to perform this search \emph{modulo commutativity}—i.e., recognize that gates may appear in any order without breaking the original functionality. Here, $g^{-1}$ is the inverse of $g$.


The current best method for handling this kind of commutativity is the Directed Acyclic Graph (DAG) based method proposed by Iten et al. \cite{iten_exact_2022}, addressed as the DAG-approach henceforth.  In the DAG-approach, the circuit is represented as a DAG, with each gate being depicted as a node, and directed edges are used to connect the gates that are interdependent and non-commutable. As a result, all circuits that are equivalent under gate commutation share identical DAG representations, which greatly facilitates template matching. However, this representation also has its issues. When two commutable gates act on the same qubits, the DAG may depict them as if they were acting on different qubits. This can undermine the connectivity and locality properties of the circuit. As a result, the matching process may require a more computationally intensive global search rather than a local one.


\begin{figure}[hbtp]
    \centering
    \includegraphics[width=0.6\linewidth]{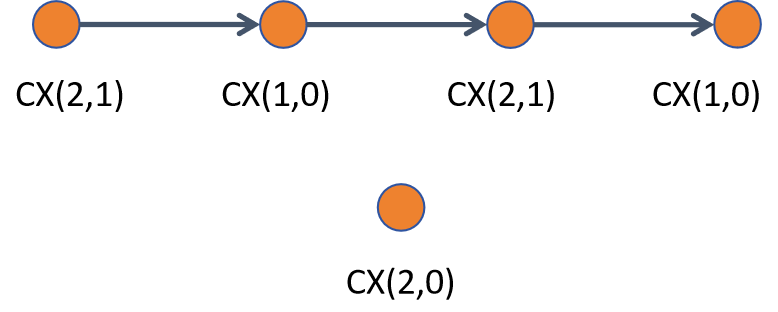}
\vspace{-10pt}
    \caption{The DAG representation of the two quantum circuits shown in Fig.~\ref{fig:qc1}. 
    }
    \label{fig:dag}
\vspace{-10pt}
\end{figure}

For example, Fig.~\ref{fig:dag} illustrates the identical DAG representation of the two circuits shown in Fig. \ref{fig:qc1}. Since $CX(2,0)$ is represented as an isolated node, we may need to traverse globally to match this gate if other gates exist, even if the four $CX$ gates ($CX(2,1);CX(1,0);$ $CX(2,1);CX(1,0)$) are matched. We lost the capacity to infer the subsequent matching gate from the previously matched gates.

\section{Hypergraph Representation
}
\label{sec:SG}

Although the DAG-approach handles gate commutativity well, it breaks the connectivity within the circuits. As a result, the matcher may waste a lot of its runtime on searching for the next matchable gate in the global region.


To mitigate this issue, in this section, we propose the hypergraph representation (HG) that merges commuting gates into a single \emph{super-node} while preserving complete connectivity and locality information.

\subsection{Basic Idea
}
\label{subsec:sg-algo}
Intuitively, the basic idea of our hypergraph representation is that we ``pinch" all the commutable nodes together. 

\begin{enumerate}
  \item \textbf{Commutation analysis.}  
  For each gate $g$, compute its eigen-directions on every qubit that $g$ acts upon.
  
  \item \textbf{Greedy merging.}  
  Iterate through the circuit. When examining a gate $g$, suppose it acts on qubits $q_1,\cdots,q_k$, and let $u_i$ be the most recently constructed super-node on $q_i$. If $g$ shares the same eigen-directions as the gates in $u_i$, then add $g$ to $u_i$; otherwise, create a new super-node $v_i$ (which contains $g$ only) and add an edge from $u_i$ to $v_i$.
  
\end{enumerate}

\subsection{Formal Definition}
\label{subsec:sg-def}

Formally, let $C$ be a quantum circuit with a set of qubits $Q$ and a set of gates $G$. Then its hypergraph representation is the hypergraph
\begin{equation}
HG(C)\,\triangleq\,(V,E),
\end{equation}
constructed as follows: 



\smallskip\noindent\textbf{Nodes.}
Each node $v\in V$ is a $3$-tuple
\begin{equation}
v=\Bigl(Q(v),\; \chi(v),\; G(v)\Bigr),
\end{equation}
where
\begin{itemize}
  \item $Q(v)\in Q$ is the qubit of node $v$;
  \item $\chi(v)$ is the eigen-directions of the quantum gates in $v$ on $Q(v)$,  called the \emph{algebraic fingerprint} of node $v$.
  \item $G(v) \subset G$ is a maximal set of gates that are adjacent and have the same eigen-directions on $Q(v)$ in $C$.
\end{itemize}
\smallskip\noindent\textbf{Edges.} 
There are two kinds of edges. First, for each multi-qubit gate $g$ in circuit $C$, we introduce a hyperedge by connecting nodes that include $g$. Second, we introduce an edge from node $u$ to node $v$ if $Q(u)=Q(v)$ and there exist  $g\in G(u)$ and $h\in G(v)$ such that $h$ is a direct successor of $g$ on qubit $Q(u)$ in circuit $C$. This means that there is a directed linear path on each qubit wire in the hypergraph.




Here are some basic properties of this graph:
\begin{itemize}
\item if $Q(u)=Q(v)$ and $u\not=v$, then $G(u)\cap G(v)=\emptyset$;
\item if gate $g\in G(u)\cap G(v)$ and $u\neq v$, then $Q(u)\not=Q(v)$ and $g$ acts on both $Q(u)$ and $Q(v)$; 
\item if $g_1$ is ahead of $g_2$ in the circuit, and $g_1 \in G(u)$ and $g_2\in G(v)$ for some $u\neq v$ and $Q(u)=Q(v)$, then $u$ is ahead of $v$.
\end{itemize}
These properties ensure that the graph corresponds to a valid quantum circuit.

\subsection{Canonicity}

In the following, we show that the hypergraph representation is canonical for circuits composed solely of multi-controlled single-target gates. This property allows it to be used for template matching without being affected by gate commutation.
\begin{thm}[Canonicity]
\label{thm:canonical}
Two circuits $C_1$ and $C_2$, consisting of only multi-controlled single-target gates, are equivalent under gate commutation iff their hypergraph representations are isomorphic:
\begin{equation}
C_1\equiv_{\text{comm}}C_2\;\Longleftrightarrow\;HG(C_1)\cong HG(C_2).
\end{equation}
\end{thm}
\begin{proof}

($\Rightarrow$). Assume $C_1 \equiv_{\text{comm}} C_2$. By definition, $C_2$ is obtained from $C_1$ via a sequence of swaps of adjacent commuting gates. It suffices to show that a single swap preserves hypergraph isomorphism. Consider two adjacent gates $g_i$ and $g_{i+1}$ in $C_1$ such that $g_i g_{i+1} = g_{i+1} g_i$. The construction algorithm processes gates sequentially; since $g_i$ and $g_{i+1}$ share identical eigen-directions on all common qubits, the super-nodes created on their common qubits are merged regardless of processing order. Nodes on disjoint qubits remain independent. Consequently, the resulting hypergraphs before and after the swap are identical, establishing $HG(C_1) \cong HG(C_2)$ by induction.

($\Leftarrow$). Assume $HG(C_1)\cong HG(C_2)$. First, the two circuits $C_1$ and $C_2$ must have the same number of qubits and gates. Consider the last gate $g$ in $C_1$. Since $HG(C_1)\cong HG(C_2)$, there must be a hyperedge in $HG(C_2)$ corresponding to the hyperedge introduced by $g$ in $HG(C_1)$. That is, there is a gate $g'$ corresponding to $g$ in $C_2$, and all nodes containing $g'$ are the last node on the corresponding qubit wire. This means $g'$ commutes with every gate that appears after it in $C_2$, allowing us to bring $g'$ to the end of the circuit via swaps of adjacent commuting gates. Consequently, we can align the final gates of $C_1$ and $C_2$. By iteratively removing these matched gates and applying the same reasoning, we can align every corresponding pair of gates through swaps of commuting gates, thereby proving $C_1 \equiv_{\text{comm}} C_2$.
\end{proof}





\vspace{-10pt}
\subsection{An Illustrative Example}

\begin{figure}[hbtp]
    \centering
\vspace{-10pt}
    \includegraphics[width=0.5\linewidth]{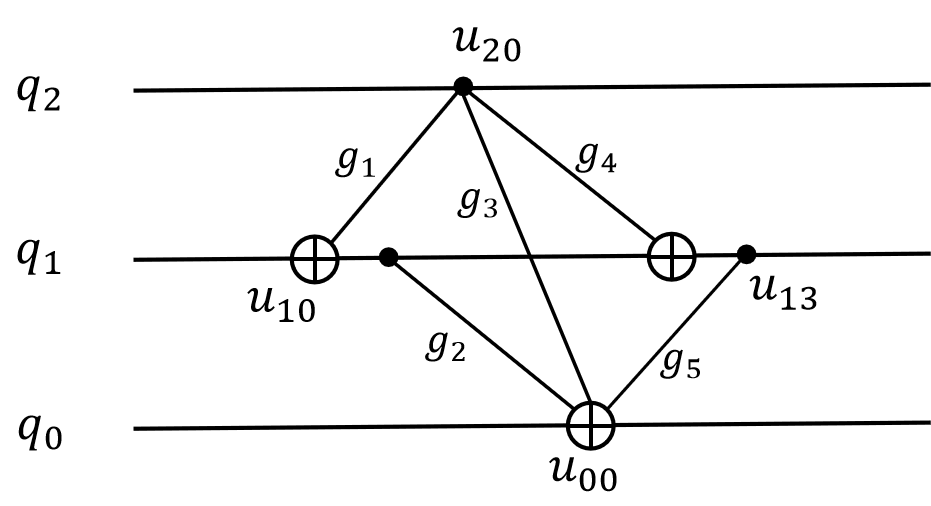}
\vspace{-10pt}
    \caption{The hypergraph representation of the quantum circuits shown in Fig. \ref{fig:qc1}. For clarity, we omitted the labels $u_{11}$ and $u_{12}$.}
\vspace{-10pt}
    \label{fig:sg}
\end{figure}

The quantum circuits shown in Fig. \ref{fig:qc1} both have the hypergraph representation as shown in Fig.~\ref{fig:sg}. This shows that two commutationally equivalent circuits yield the \emph{identical} hypergraph representation, validating Theorem~\ref{thm:canonical}.


Let's illustrate the construction process using the circuit shown in Fig.~\ref{fig:qc1}~(a) as an example. First, we introduce three qubit wires labeled \( q_2 \), \( q_1 \), and \( q_0 \), corresponding to the three qubits in the original circuit. Next, we examine the first gate \( g_1 (CX(2,1)) \). For this gate, we create two nodes, \( u_{20} \) and \( u_{10} \), and connect them with an edge. Specifically, \( Q(u_{20}) = q_2 \), \( G(u_{20}) = \{g_1\} \), and \( \chi(u_{20}) = \{\ket{0}, \ket{1}\} \), while \( Q(u_{10}) = q_1 \), \( G(u_{10}) = \{g_1\} \), and \( \chi(u_{10}) = \{\ket{+}, \ket{-}\} \). For simplicity, we represent \( u_{20} \) as \( \cdot \) and \( u_{10} \) as \( \oplus \), analogous to their depictions in quantum circuits. Similarly, for the gate \( g_2 (CX(1,0)) \), we construct two nodes \( u_{11} \) and \( u_{00} \). When processing the third gate \( g_3 (CX(2,1)) \), we create only one new node \( u_{12} \), merging the other node into \( u_{20} \), which updates \( G(u_{20}) \) to \( \{g_1, g_4\} \). This process continues iteratively until all gates are processed, ultimately yielding the hypergraph shown in Fig.~\ref{fig:sg}. The same hypergraph can also be constructed for the circuit shown in Fig.~\ref{fig:qc1}~(b), which is commutationally equivalent to that in Fig.~\ref{fig:qc1}~(a).


\subsection{Comparison with the DAG-approach}
Let $C$ be a quantum circuit and $T$ be a template. 
Note that the DAG-approach needs to search globally to find the optimal matching, while only local search is needed for the HG-based method. 
A comparison between HG and DAG is shown in Table \ref{tab:com_dag}.

\begin{table}[h]
\centering
\caption{Complexity comparison with DAG-based matcher~\cite{iten_exact_2022}.}
\vspace{-10pt}
\resizebox{\columnwidth}{!}{%
\begin{tabular}{lcc}
\hline
 & DAG~\cite{iten_exact_2022} & HG\\
\hline
Construction & $O(n_C|C|^2)$ & $O(n_C|C|)$ \\
Commute & \checkmark &\checkmark \\
Connectivity &\ding{53} &\checkmark \\
Locality &\ding{53} &\checkmark\\
Complexity &$O(|C|^{|T|+3}|T|^{|T|+4}n_C^{n_T-1})$ & $O(|C|^{|T|+1}|T|^{|T|}n_C^2)$\\
Matching speed & baseline & $\mathbf{11\times\text{-- } 606\times}$ \\
\hline
\end{tabular}}
\vspace{-10pt}
\label{tab:com_dag}
\end{table}

The hypergraph representation retains the commutativity-aware efficiency of the DAG-approach while recovering \emph{all} gate-level details necessary for semantic validation and local search.
By reducing the search space and eliminating spurious candidates early, it forms the core of our accelerated template-matching engine described in Section~\ref{sec:match}. The complexity for constructing the hypergraph representation is $O(n_C|C|)$ for a circuit $C$, since $O(k)$ steps are needed for coping with a $k$ qubit gate. The complexity of the HG-based matcher will be presented in Section~\ref{sec:complexity}.

\section{Template Matching via Hypergraphs
}
\label{sec:match}

We now describe how the hypergraph representation (HG) is used to accelerate template matching.
The algorithm takes as input a target circuit $C$ and a template $T$.
It searches for all best subgraphs of $HG(C)$ that are isomorphic to a connected subgraph of $HG(T)$, and replaces the matched gates in $C$ whenever the matched part contains more than $|T|/2$ gates. The resulting circuit is guaranteed to be equivalent to $C$ and contains normally fewer gates. 

We use an exhaustive depth-first search method to find the optimal match. What should be noticed here is that the connectivity of the matched gates should be checked, so that they can be arranged into a valid quantum circuit.
The pseudocode of our algorithm can be seen in Algorithm \ref{alg:tm}. Throughout the search,  we maintain and incrementally update three partial mappings to record the matched nodes, qubits, and gates:
\begin{itemize}
    \item \textit{$M$ (matched nodes)}: $V(T) \rightharpoonup V(C)$,
    \item \textit{$M_q$ (matched qubits)}: $Q(T) \rightharpoonup Q(C)$,
    \item \textit{$M_g$ (matched gates)}: $G(T) \rightharpoonup G(C)$.
\end{itemize}

\begin{algorithm}
\caption{Hypergraph Template Matcher}

\begin{algorithmic}[1]

\Require{A target circuit $C$, and a template $T$}
\Ensure{An equivalent circuit with the same number or fewer gates}

    \ForAll{node $v \in V(HG(C))$}
    \State $(M^m, M_q^m, M_g^m) \gets (\emptyset, \emptyset, \emptyset)$ 
        \ForAll{node $u \in V(HG(T))$ with $Q(u)=Q(v)$ and $\chi(u)=\chi(v)$}
            \State $(M, M_q, M_g) \gets$ \textsc{GateMatch}$(\{u:v\}, \{Q(u):Q(v)\}, \emptyset)$
            \If{$|M_g| > |M_g^m|$}
                \State $(M^m, M_q^m, M_g^m) \gets (M, M_q, M_g)$ 
            \EndIf
        \EndFor
        
            \If{$|M_g^m| > |T|/2$}
                \State rem $\gets$ InverseRemainder$(T,M_g^m)$
                \State $C \gets$ Replace$(C,M_g^m,rem)$
            \EndIf
        
    \EndFor
\State \Return $C$
\\
\Function{GateMatch}{$M,M_q,M_g$}
    \State $(M^t, M_q^t, M_g^t) \gets (\emptyset, \emptyset, \emptyset)$ 
    \ForAll{gate $g \in T$ not separated from matched gates and not yet matched}
        \ForAll{gate $g' \in C$ with same unitary and topology}
                \State $M_g \gets M_g \cup \{g:g'\}$
                \State Update $M,M_q$
                \State $(M, M_q, M_g) \gets$ \textsc{GateMatch}($M,M_q,M_g$)
                \If{$|M_g| > |M_g^t|$}
                \State $(M^t, M_q^t, M_g^t) \gets (M, M_q, M_g)$ 
                \EndIf
        \EndFor
    \EndFor
    \State \Return $(M^t, M_q^t, M_g^t)$
\EndFunction

\end{algorithmic}

\label{alg:tm}

\end{algorithm}

\subsection{Initial Node Match}


Let $v$ be a node in $HG(C)$. We first consider finding a matched subgraph in the vicinity of $v$.

The process starts by selecting a node $u\in V(HG(T))$ such that $\chi(u) = \chi(v)$.
Then, we initialize
\[
\textit{$M$}\gets\{u:v\},
\]
\[
\textit{$M_q$}\gets\{Q(u):Q(v)\}
\]
and grow the isomorphism starting from this pair.

\subsection{Next Gate Match}

At each step, we pick an unmatched gate $g\in T$ and try to pair it with a gate $g'\in C$ that
\begin{enumerate}
  \item implements the same unitary (e.g., both are CX), and
  \item satisfies the three feasibility rules:
\end{enumerate}
\begin{itemize}
    \item \textbf{Node consistency}: the new pair does not cause conflict with matched nodes;
    \item \textbf{Qubit consistency}: the new pair does not cause conflict with matched qubits;
    \item \textbf{Gate adjacency}: the newly matched gate is not separated from already matched gates by any unmatched gate.
\end{itemize}
The third rule ensures that all matched gates form a valid subcircuit (up to gate commutation) of the original circuit.

Whenever the rules hold, we add $\{g:g'\}$ to $M_g$ and extend $M$ and $M_q$ correspondingly, and continue recursively. More detailed explanations can be found in Section \ref{sec:example}, illustrated with an example.

\subsection{Exhaustive Traversal}

Note that many different choices can be made during every step, including initial node match and next gate match. Thus, we traverse over all possible choices and find the optimal one.
This exhaustive strategy ensures that the largest possible isomorphic subgraph is found for the starting node $v$.


\subsection{Equivalent Replacement}

Once the optimal match is obtained, we check whether the number of matched gates exceeds $|T|/2$.
If so, we construct the inverse remainder or the matched gates in $T$. We calculate the inverse of the unmatched gates. It is easy to distinguish whether each gate is in front of or behind the already matched segments, thereby determining their positions in the inverse remainder. The matched segment in $C$ is then replaced by the inverse remainder, yielding a circuit that is strictly smaller and functionally equivalent. 


\subsection{Complexity}
\label{sec:complexity} 

Let $C$ be the target circuit and $T$ be the template. Recall that we use $n_C$ and $n_T$ to denote the number of qubits, $|C|$ and $|T|$ the number of gates. 

For each node $v \in HG(C)$, the initial node matching complexity is $O(n_T|T|)$. The complexity of matching the first gate is $O(|T||C|)$. From now on, for each subsequent match, only the remaining gates need to be considered. However, each time a new gate is matched, a separation test needs to be conducted, with the complexity being $O(n_T|T|+n_C|C|)$. Normally, $n_T|T|<n_C|C|$.

Therefore, the total complexity is taken as $O(n_Tn_C|C||C|^{\underline{|T|}}|T|!)$, less than $O(|C|^{|T|+1}|T|^{|T|}n_C^2)$. Here, $|C|^{\underline{|T|}}=|C|\cdot(|C|-1)\cdots(|C|-|T|+1)$.


\section{An Example}\label{sec:example}

\begin{figure}[htbp]
\vspace{-20pt}
\centering
\subfigure[]{
\begin{tikzpicture}
  \begin{yquant}[register/minimum height=7mm, operator/separation=1.6mm, control style={radius=1.7pt}, subcircuit box style={dashed}]
    qubit {$q_2$} q2;
    qubit {$q_1$} q1;
    qubit {$q_0$} q0;

    cnot q1 | q2;
    cnot q0 | q1;
    box {$Y$} q1 | q2;
    cnot q0 | q1;
    cnot q0 | q2;
    
  \end{yquant}
  
\end{tikzpicture}
}
\subfigure[]{
\includegraphics[width=0.48\linewidth]{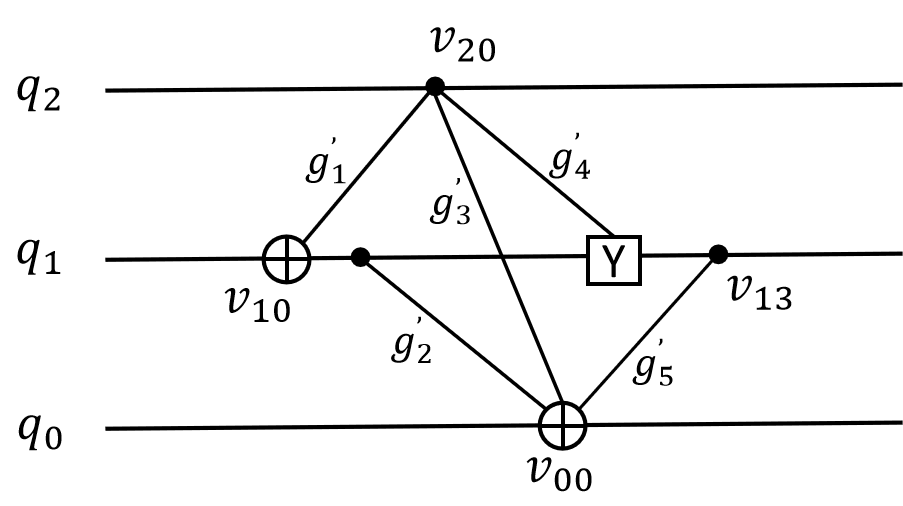}
}
\vspace{-10pt}
\caption{A quantum circuit with 3 qubits and 5 gates and its hypergraph representation. We omitted the labels $v_{11}$ and $v_{12}$ here.}
\vspace{-10pt}
\label{fig:qc3}
\end{figure}

We now demonstrate the matching algorithm through a concrete example. Consider the target circuit depicted in Fig.~\ref{fig:qc3} (a) and its hypergraph representation in Fig.~\ref{fig:qc3} (b). We aim to simplify this circuit using the template from Fig.~\ref{fig:qc1} (with its hypergraph shown in Fig.~\ref{fig:sg}).


\textbf{Initialization.} We begin by examining each node in the target hypergraph (Fig.~\ref{fig:qc3}), selecting the first $\oplus$-node on qubit $q_1$ (i.e., node $v_{10}$) as our starting point. Within the template hypergraph, candidate matches for $v_{10}$ include either of the two $\oplus$-nodes on $q_1$ or the $\oplus$-node on $q_0$. Suppose we match $v_{10}$ with the first $\oplus$-node on $q_1$ in the template hypergraph. This yields the initial mappings: $M = \{u_{10} : v_{10}\}$ and $M_q = \{q_1 : q_1\}$.


\textbf{First Gate Matching.} Both $G(u_{10})$ and $G(v_{10})$ contain exactly one gate ($g_1$ and $g'_1$, respectively) with identical functionality, enabling their direct pairing. The mappings are consequently updated to: $M = \{u_{10}:v_{10}, u_{20}:v_{20}\}$, $M_q = \{q_1:q_1, q_2:q_2\}$, and $M_g = \{g_1:g_1'\}$. 


\textbf{Subsequent Gate Exploration.} Proceeding to the vicinity of matched nodes, we observe that $u_{20}$ contains two unmatched gates, $g_3$ and $g_4$. Since both are $\mathrm{CX}$ gates, they can only pair with the corresponding $\mathrm{CX}$ gate $g'_3$ in node $v_{20}$ (matching $g_4$ with $g'_3$ would violate qubit consistency constraints in $M_q$). Thus, we establish the mapping $g_3 : g'_3$, resulting in: $M = \{u_{10}:v_{10}, u_{20}:v_{20},u_{00}:v_{00}\}$, $M_q = \{q_1:q_1, q_2:q_2, q_0;q_0\}$, and $M_g = \{g_1:g_1',g_3:g_3'\}$.


\textbf{Constraint Validation.} We next evaluate the remaining gates $g_2$ and $g_5$ in $u_{00}$. Matching $g_2$ with $g'_5$ would introduce an intervening gate $g'_4$ that separates $g'_5$ from previously matched gates, violating the gate adjacency feasibility rule (i.e., the matched subgraph would not correspond to a valid subcircuit). Consequently, this match is discarded. The pairing of $g_2$ with $g'_2$ satisfies all constraints, whereas matching $g_5$ with either $g'_2$ or $g'_5$ would similarly violate subcircuit validity.

\textbf{Result Summary.} Starting from $M = \{u_{10} : v_{10}\}$, we successfully match three gates: $\{g_1 : g'_1, g_3 : g'_3, g_2 : g'_2\}$. These can be replaced by the two remaining gates from the template, achieving a net reduction of one gate.


Alternative starting points yield suboptimal matches:
\begin{itemize}
    \item From $M = \{u_{00} : v_{10}\}$: at most two gates ($\{g_3 : g'_1, g_1 : g'_3\}$);
    \item From $M = \{u_{12} : v_{10}\}$: at most two gates ($\{g_4 : g'_1, g_3 : g'_3\}$).
\end{itemize}


Throughout this process, the search is confined exclusively to matched nodes and their immediate neighborhoods, eliminating the need to examine gates on other qubits or distant circuit regions. This localized exploration exemplifies the computational advantage of our hypergraph approach over DAG-based methods.


\section{Experiments}\label{sec:experiments}

We compare our HG-based method with the state-of-the-art DAG-based matcher~\cite{iten_exact_2022} implemented in Qiskit \cite{javadi2024quantum}. The DAG-approach is divided into two versions. One of them incorporates a heuristic, which reduces the matching time at the cost of sacrificing some optimization effects. All experiments run on an Intel i7-12700H CPU with 32 GB RAM; our matcher is implemented using the Python programming language, the same as Qiskit. Times are wall-clock seconds.

\subsection{Random Circuits}
\label{subsec:random}

Following~\cite{iten_exact_2022}, we first generate random circuits with gate set \{X, CX, CCX\} on 10 qubits. Circuit sizes are 100, 300, 500, and 700 gates. The template library contains 24 templates with $\le 8$ gates (Qiskit's “nct” templates excluding the 9-gate ones). Every optimizer is invoked \emph{once} per circuit. If we include the 9-qubit gate templates, the time required by the DAG-based method will increase significantly. In such a case, the gap between our method and theirs will further widen.

\begin{table}[h]
\centering
\vspace{-10pt}
\caption{Random-circuit reduction (10 qubits).}
\label{tab:random}
\vspace{-10pt}
\resizebox{0.9\columnwidth}{!}{%
\begin{tabular}{lrrrrrr}
\hline
 & \multicolumn{2}{c}{HG} & \multicolumn{2}{c}{Qiskit} & \multicolumn{2}{c}{Qiskit-heur} \\
\cline{2-3}\cline{4-5}\cline{6-7}
\#gates & \#gates & time (s) & \#gates & time (s) & \#gates & time (s) \\
\hline
100 & 73 & 1.75 & 73 & 27.64 & 75 & 18.97 \\
300 & 226 & 5.67 & 226 & 177.50 & 228 & 93.20 \\
500 & 390 & 9.19 & 390 & 583.34 & 393 & 293.85 \\
700 & 550 & 20.26 & 549 & 1321.91 & 556 & 587.15 \\
\hline
\end{tabular}%
}
\vspace{-10pt}
\end{table}

The experimental data are recorded in Table~\ref{tab:random}, which includes the number of gates before and after optimization, as well as the time required by different optimizers. The table shows that HG obtains very similar gate counts while being \textbf{15--65$\times$ faster} than Qiskit and \textbf{11--29$\times$ faster} than Qiskit with heuristics enabled.

\subsection{Fixed Benchmarks}
\label{subsec:fixed}

We next repeat the protocol of \cite{iten_exact_2022}: take the circuits already optimized by \cite{nam2018automated} and apply template matching until the gate count stabilizes. Results for the arithmetic suite are listed in Table~\ref{tab:fixed}. The HG matcher stops at the same fixed point as Qiskit but uses \textbf{two orders of magnitude less time}. On the largest instance \texttt{gf2\string^E16\_mult} (2707 initial gates, 48 qubits) HG finishes in 21 s versus 12 737 s for Qiskit, a 606$\times$ speed up.

\begin{table}[ht]
\centering
\vspace{-10pt}
\caption{Arithmetic circuits from \cite{nam2018automated}; repeated matching until fixed.}
\label{tab:fixed}
\vspace{-10pt}
\resizebox{0.9\columnwidth}{!}{%
\begin{tabular}{lrrrrrr}
\hline
 & & & \multicolumn{2}{c}{Qiskit} & \multicolumn{2}{c}{HG} \\
\cline{4-5}\cline{6-7}
circuit & \#gates & \#qubits & \#gates & time (s) & \#gates & time (s) \\
\hline
adder\_8 & 606 & 24 & 601 & 91.00 & 598 & 1.54 \\
csla\_mux\_3 & 155 & 15 & 153 & 8.98 & 153 & 0.26 \\
csum\_mux\_9 & 266 & 30 & 266 & 11.75 & 266 & 0.27 \\
gf2\string^E10\_mult & 1070 & 30 & 1033 & 1322.18 & 1033 & 9.95 \\
gf2\string^E16\_mult & 2707 & 48 & 2617 & 12737.50 & 2599 & 20.99 \\
gf2\string^E4\_mult & 187 & 12 & 183 & 11.30 & 183 & 0.69 \\
gf2\string^E5\_mult & 296 & 15 & 289 & 28.26 & 289 & 1.36 \\
gf2\string^E6\_mult & 403 & 18 & 392 & 67.16 & 392 & 2.32 \\
gf2\string^E7\_mult & 555 & 21 & 539 & 216.33 & 539 & 3.85 \\
gf2\string^E8\_mult & 712 & 24 & 688 & 360.02 & 687 & 6.10 \\
gf2\string^E9\_mult & 891 & 27 & 864 & 593.61 & 862 & 8.59 \\
mod5\_4 & 51 & 5 & 49 & 1.78 & 49 & 0.12 \\
mod\_mult\_55 & 91 & 9 & 91 & 1.70 & 91 & 0.09 \\
mod\_red\_21 & 180 & 11 & 180 & 4.00 & 180 & 0.16 \\
qcla\_adder\_10 & 399 & 36 & 393 & 143.27 & 393 & 0.91 \\
qcla\_com\_7 & 284 & 24 & 280 & 31.28 & 280 & 0.65 \\
qcla\_mod\_7 & 624 & 26 & 613 & 207.17 & 612 & 1.67 \\
rc\_adder\_6 & 140 & 14 & 140 & 2.70 & 140 & 0.12 \\
toff\_barenco\_10 & 264 & 19 & 264 & 12.85 & 264 & 0.21 \\
toff\_barenco\_3 & 40 & 5 & 40 & 0.50 & 40 & 0.03 \\
toff\_barenco\_4 & 72 & 7 & 72 & 1.29 & 72 & 0.05 \\
toff\_barenco\_5 & 104 & 9 & 104 & 2.37 & 104 & 0.08 \\
toff\_nc\_10 & 175 & 19 & 175 & 4.21 & 175 & 0.12 \\
toff\_nc\_3 & 35 & 5 & 35 & 0.39 & 35 & 0.02 \\
toff\_nc\_4 & 55 & 7 & 55 & 0.63 & 55 & 0.04 \\
toff\_nc\_5 & 75 & 9 & 75 & 1.02 & 75 & 0.04 \\
vbe\_adder\_3 & 89 & 10 & 86 & 3.20 & 82 & 0.23 \\
\hline
\end{tabular}%
}
\vspace{-12pt}
\end{table}

\subsection{Scalability}
\label{subsec:scale}

\begin{figure}[h]
\centering
\includegraphics[width=0.85\linewidth]{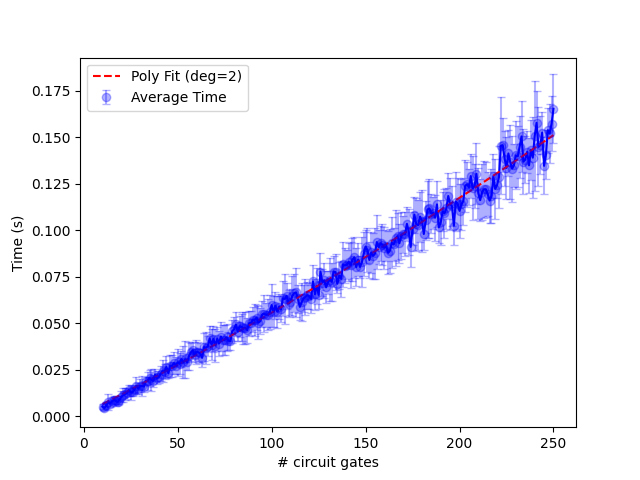}
\vspace{-10pt}
\caption{Average optimisation time for 6-qubit random circuits (15 runs per point). Error bars show standard deviation.}
\label{fig:scale}
\vspace{-10pt}
\end{figure}

We finally reproduce the scaling experiment of \cite{iten_exact_2022}: 6-qubit random circuits with gate counts from 10 to 250 (15 instances per point). Fig.~\ref{fig:scale} shows average optimisation time versus circuit size. A quadratic fit
\[
y = 3.75\!\times\!10^{-7}\,x^{2} + 5.04\!\times\!10^{-4}\,x + 1.55\!\times\!10^{-3}
\]
is essentially linear in the observed range, confirming that HG scales near-linearly and remains $<200$ ms even for the largest circuits. As a comparison, the quadratic fit for DAG-based method is (taken from~\cite{iten_exact_2022})
\[
y = 6.8\!\times\!10^{-5}\,x^{2} + 1.26\!\times\!10^{-3}\,x + 8.84\!\times\!10^{-3}.
\]

\subsection{Summary}
Across all three test suites, HG consistently matches the gate counts produced by Qiskit while delivering \textbf{11--606$\times$ speed-up}. The acceleration grows with circuit size, validating the efficiency of the hypergraph approach.

\section{Conclusion}\label{sec:conclusion}

We have presented the \emph{hypergraph representation}, a commutativity-aware representation that collapses commuting gates into super-nodes while preserving complete gate-level information.
By coupling this representation with a lightweight subgraph isomorphism search, our hypergraph representation matcher achieves \textbf{11--606$\times$ speed-up} over the state-of-the-art DAG-based implementation in Qiskit on standard benchmark suites while maintaining the same optimization quality.

Extensive experiments on random circuits, fixed arithmetic benchmarks, and scalability tests demonstrate that the acceleration grows with circuit size, confirming the near-linear scalability of the HG approach. 



\begin{acks}
Work partially supported by Quantum Science and Technology-National Science and Technology Major Project under Grant No. 2024ZD0300502,  Beijing Nova Program Grant No. 20240484652, Beijing Municipal Science and Technology Program Grant No. Z25110100810000.
\end{acks}

\clearpage
\bibliographystyle{ACM-Reference-Format}
\bibliography{reference}

\end{document}